\documentclass[fleqn,usenatbib]{mnras}

\usepackage{newtxtext,newtxmath}

\usepackage[T1]{fontenc}

\DeclareRobustCommand{\VAN}[3]{#2}
\let\VANthebibliography\thebibliography
\def\thebibliography{\DeclareRobustCommand{\VAN}[3]{##3}\VANthebibliography}

\usepackage{graphicx}	
\usepackage{amsmath}	

\title[Spectroscopic Binaries]{Search for spectroscopic binaries using rotational velocities in five open clusters observed by ESO.}

\author[M. Kovalev et al.]{
Mikhail Kovalev,$^{1,2,3,4}$ \thanks{E-mail: mikhail.kovalev@ynao.ac.cn}
Mariyam Ahmed,$^{5}$
Randa Asa'd,$^{5}$ \thanks{E-mail: raasad@aus.edu}
\\
$^{1}$Yunnan Observatories, China Academy of Sciences, Kunming 650216, China\\
$^{2}$Key Laboratory for the Structure and Evolution of Celestial Objects, Chinese Academy of Sciences, Kunming 650011, China\\
$^{3}$Sternberg Astronomical Institute, Leninskie Gory, Moscow 119992, Russia\\
$^{4}$International Centre of Supernovae, Yunnan Key Laboratory, Kunming 650216, China\\
$^{5}$Physics Department, American University of Sharjah, PO Box 26666, Sharjah, UAE\\
}

\date{Accepted XXX. Received YYY; in original form ZZZ}

\pubyear{2023}

\def\kms{\,{\rm km}\,{\rm s}^{-1}}

\def\feh{\hbox{[Fe/H]}}

\newcommand{\teff}{T_{\rm eff}}
\newcommand{\rv}{{\rm RV}}
\def\Vmic{V_{\rm mic}}

\def\vsini{(V \sin{i})}
\def\logg{\log{\rm (g)}}
\def\snr{\hbox{S/N}}
\newcommand{\ha}{\hbox{H$\alpha$}}

\begin{document}
\label{firstpage}
\pagerange{\pageref{firstpage}--\pageref{lastpage}}
\maketitle

\begin{abstract}
In this paper we detect double-lined spectroscopic binaries (SB2) in five open clusters: NGC 2243, NGC 2420, NGC 3532, NGC 6253 and NGC 6705 (M 11) using a method based on high values of the projected rotational velocity when they are fitted with single star spectral model. Observed spectra were obtained from ESO archive.  The method was validated on sets of synthetic spectra for the single and binary stars. It is able to reliably select spectroscopic binaries without confusing them with single stars, if components in binary rotate slowly and radial velocity separation is sufficiently high. We found 60 SB2 candidates: two in NGC~2243, eight in NGC~2420 and NGC~3532, 17 in NGC 6253 and 25 in NGC~6705. Comparison with literature confirms 18 of them, thus we found 42 new SB2 candidates. 

\end{abstract}




\section{Introduction}

Binary systems are a major area of astronomical study for several reasons. Approximately 40\% of stars in the  Milky Way are part of a binary or higher-order system \citep{Moe_2017}. Unidentified binary spectra are also potentially responsible for errors in atmospheric parameter estimation performed by data processing pipelines associated with spectroscopic surveys, thus cataloguing binaries can improve the accuracy of these estimations \citep{intr1,elbadry1}. In addition to being a direct method of estimating stellar masses by using relations derived from Kepler’s laws \citep{intr2},  detecting binary systems and studying their orbital elements allows us to constrain models of stellar evolution. Spectra from binary stars can be used to determine the properties of each star in the system, such as its mass, size, temperature, and luminosity. This information can then be used to gain insight into the processes that govern the formation and evolution of stars. How the stars in a binary system interact with each other, exchange mass, and evolve over time can be understood by observing binary systems at different stages of their evolution. This observation can provide important clues about the processes that govern the evolution of stars and the formation of new stars. \\
Binary systems are of three main types, mainly divided by the method used to observe and study their components and dynamics. A visual binary is one that can be resolved by telescopes into two separate objects. Such binaries provide the advantage of allowing the measurement of all orbital elements associated with it, since their orbits can be traced by observing the stars’ movement over time \citep{intr3}. However, the detection of visual binaries is subject to the limitation of distance as it is constrained by the  angular resolution of the instrument used. The second group of binary stars are eclipsing binaries, where the orbit of the system is edge-on to the line of sight. These binaries are instead studied by measuring the dip in flux that occurs when one component passes in front of or behind the other.\\
Finally, the third kind of binaries are spectroscopic binaries (SB). These stars are too close together to be resolved as distinct objects and thus appear as one star. Despite it being not possible to resolve the two components, the presence of second star can be inferred from peculiarities in the spectra recorded from the system. The spectra of an SB system also reveal the stars' motion relative to one another. As the two stars orbit each other, they periodically reach points in their orbit when they are moving away from or towards Earth. However, because the orbit of the system is not always perpendicular to us, we only observe the effect of the component of the movement that is in our direction. Consequently, when some component of their orbital velocities is in our line-of-sight, their observed spectral lines shift back and forth due to the Doppler effect.  SB systems are particularly useful for studying the properties of binary stars, as they provide an opportunity to observe the properties of two stars over much longer distances than visual binaries. This fact allows SB systems to even be detected in even other nearby galaxies \citep{2020A&A...635A.155M}. These three categories are not mutually exclusive; eclipsing binaries are special cases of visual or, most commonly, spectroscopic binaries.\\
SBs are further classified by the number of spectral components, such as spectral lines, that are visible in their spectra (SBn, where n $\geq$ 1). In SB1 stars, the spectrum of one star dominates the other due to the primary star being much brighter, and only one spectrum is visible. These star systems are classified as binaries by detecting periodic shifts in the radial velocity of the primary component that cannot be explained by other reasons such as pulsations. The detection of SB2s is simpler since the spectral components of both stars are clearly visible when a component of their orbital movement is in our line-of-sight. The consequent Doppler shift of each of their spectra causes a separation of their measured radial velocities. SB2s with non-zero radial velocity separation of components are usually detected by analysis using cross-correlation functions (CCF) and detecting of multiple peaks in the CCF \citep{2017A&A...608A..95M}. In contrast, machine-learning algorithms are also able to detect SB2s with |$\Delta$ RV| close to zero by fitting the observed spectrum to composite spectral models \citep{elbadry1,elbadry2,2022MNRAS.510.1515K}. \\
 However detecting SB2 is not straight forward, as composite spectral model may not be picking up the contribution of a secondary star, but may just be fitting to the spectral noise. To further improve selection criteria to reliably detect SB2 candidates we can use projected rotational velocity $\vsini$. This is the speed at which star rotates around it's axis. Stellar rotation is one factor that can cause the broadening of spectral lines, another being the presence of a binary system where the two spectral components are too close to be resolved separately. \cite{2022MNRAS.517..356K} used a new selection criterion to find SB2s from spectral analysis, which involves the projected rotational velocity estimated for the single and binary spectral fit. Thus, this selection criteria works under the principle of separating rapidly rotating single stars, with broad spectral components, from two slowly rotating stars in a binary system. \cite{2022MNRAS.517..356K} tested this new method on observations from the NGC 6705 (M11) cluster and LAMOST-MRS data \citep{2020arXiv200507210L}.\\
The current paper aims to validate this selection method on five different open clusters NGC 2243, NGC 2420, NGC 3532, NGC 6253 and NGC 6705. These clusters vary significantly in both age and metallicity, so they are good test subjects to validate the reliability of the method.
 
\section{Data}
 In this study we arbitrary select five open star clusters in the European Southern Observatory (ESO) archive to realistically detect SB2 stars. The spectra in all clusters were obtained with the VLT (Very Large Telescope))/GIRAFFE spectrograph \citep{2002Msngr.110....1P} using the HR15N setup, which spans 380 Å, from 6440 Å to 6820 Å, at a resolution of \emph{R} = $\lambda$/$\Delta$$\lambda$ $\approx$ 19200. This setting includes the $\ha$ line, a deep spectral line of the hydrogen atom with a wavelength of approximately 6563 Å, to allow for easier distinction between the two spectral components of SB2s. Additionally there are more spectral observations in HR15N setting than in others \citep{2022A&A...666A.121R}.
\par
 The selected clusters are listed in Table~\ref{tab:sample}, which listed literature age and metallicities together with the number of available spectra selected inside one degree cone. All spectra are publicly available in ESO portal\footnote{\url{http://archive.eso.org/wdb/wdb/adp/phase3_spectral/form}}  and were observed as a part of Gaia-ESO survey \citep{Gilmore2012}, except for NGC 6253 which was observed earlier. Each target has only one observation in HR15N setting.  We use all available spectra and did not check their membership for each cluster. 

\begin{table}
    \centering
    \begin{tabular}{lccc}
\hline
open cluster & $\log{\rm Age}$, yrs & $\feh$, dex & $N_{\rm spectra}$\\
\hline
NGC~2243 & 9.64 & $-0.45\pm0.05$ & 640 \\
NGC~2420 & 9.24 & $-0.15\pm0.02$ & 542 \\
NGC~3532 & 8.6 & $-0.03\pm0.08$ & 906 \\
NGC~6253 & 9.54 & 0.30 & 202 \\
NGC~6705 (M~11)& 8.49 & $0.03\pm0.05$ & 1025 \\
\hline

    \end{tabular}
    \caption{The list of selected open clusters. Ages and metallicities are from \protect\cite{2022A&A...666A.121R}, except for NGC~6253 with values from \protect\cite{2010A&A...509A..17D}.}
    \label{tab:sample}
\end{table}

\section{Method}
We use the method introduced in \cite{2022MNRAS.517..356K} to detect SB2 stars. The observed spectra are fit to a single star and binary system synthetic model, and the associated goodness-of-fit measure, the reduced $\chi^2$ score, is estimated for both.
\begin{equation}
    \chi^2=\frac{1}{N-p} \sum_{\lambda=\lambda_{min}}^{\lambda=\lambda_{max}} \left[\frac{f_{\lambda,{\rm obs}}-f_{\lambda,{\rm model}}}{\sigma_\lambda}\right]^2
	\label{eq:chisq}
\end{equation}

where $N=7441$ is the number of wavelength points in the HR15N observation, $p$ is the number of estimated parameters for a single star or binary model, and $\sigma_\lambda$ is the uncertainty (error) of the observed flux.  Each model is also multiplied by a linear combination of first four Chebyshev polynomials, which is used for normalisation, similarly to \cite{2022MNRAS.517..356K}. 
We used simple linear interpolation\footnote{ with \sc{scipy.interpolate.LinearNDinterpolate}} in the grid of 1083 spectral models, computed using NLTE~MPIA spectral synthesis interface\footnote{\url{https://nlte.mpia.de}} \citep[see Chapter~4 in][ for details]{disser} to generate single-star stellar model, using {\sc SIU} radiative transfer code by \cite{siu}. We used MAFAGS-OS atmospheric models \citep{Grupp2004a,Grupp2004b}.
The grid parameters include the effective temperature ($\teff$), surface gravity ($\logg$), metallicity ([Fe/H]) and projected rotational velocity $\vsini$ randomly selected within the following ranges $\teff=5080,8720$ K, $\logg=2.9,4.7$ cgs, $\feh=-0.46,0.45$ dex and $\vsini=5,290\, \kms$. These ranges were selected to insure stable work of the linear interpolation. Microturbulence was set to $\Vmic=2\,\kms$  for all models, for simplicity. We also fit for a radial velocity, therefore single star model is characterised by $p=9$ parameters.  
\par
For the binary system model, the composite spectrum is generated by summing two single star spectral models which are scaled according to the relative luminosities of each star. The luminosity of a star is a function of its effective temperature and stellar size.

The flux of the binary model is thus given by 
\begin{equation}
    f_{\lambda,{\rm binary}}  = \frac{f_{\lambda,1}  + k_\lambda  f_{\lambda,2}}{(1 + k_{\lambda})},\\
    k_{\lambda}=\frac{B_\lambda({\teff}_1)M_1}{B_\lambda({\teff}_2)M_2} (\logg_2-\logg_1)
    \label{eq:binaryflux}
\end{equation}

where $k_{\lambda}$ is the luminosity ratio of the two binary stars per wavelength unit,  computed using black body radiation $B_\lambda(\teff)$(Plank formula) and ratio of radii, derived using mass ratio and surface gravities. Mass ratio $q=M_1/M_2$ is fitted in the range $q=0.1,\,10$. Metallicity is fixed to the same value for both components. We also fit for radial velocities of both components, therefore binary star model is characterised by $p=14$ parameters.
\par
Once the observation is fitted to both single and binary models, the improvement factor, ${\rm imp}$, is calculated to compare both fits:
\begin{equation}
    {\rm imp} = \frac{\sum_{\lambda=\lambda_{min}}^{\lambda=\lambda_{max}}(|f_{\rm \lambda,single}- f_{\rm \lambda,obs}|- |f_{\rm \lambda,binary}-f_{\rm \lambda,obs}|)/\sigma_\lambda}{\sum_{\lambda=\lambda_{min}}^{\lambda=\lambda_{max}}(f_{\rm \lambda,single}-f_{\rm \lambda,binary})/\sigma_\lambda}
    \label{eq:fimp}
\end{equation}

where $\sigma_\lambda$ is the uncertainty (error) associated with the observed flux. The improvement factor estimates how much the binary model improves the fit compared to the single model.

\section{Selection Criteria}

Binary candidates are selected using the empirical criterion from  \citep{2022MNRAS.517..356K}:
\begin{equation}
    \vsini_1+\vsini_2 +\vsini_{\rm min}< \vsini_0      
    \label{eq: selection}
\end{equation}
where $\vsini$ represents the projected rotational velocity,  with subscripts $0, 1, 2$ for the single star model, the primary/secondary components of the binary model, respectively and $\vsini_{\rm min}$ takes to account possible uncertainty in measurements. This selection criterion is a new method proposed by \cite{2022MNRAS.517..356K} for selecting SB2s and was inspired by the degeneracy between $|\rv_1 - \rv_2|$ and $\vsini_0$ of SB2s. The rationale for this criterion can be understood by considering the possibility of fitting a true single star with a binary model and vice versa:
\begin{enumerate}
    \item Fitting a true single star with binary model: the first possibility is the binary model selected will consist of two spectra with minimal RV separation, resembling twin stars. Thus, the estimated rotational velocities $\vsini_{1,2} \sim \vsini_0$, or $2 \vsini_0 = \vsini_1 + \vsini_2$. It is thus evident that most single stars will follow this relationship. On the other hand, it is possible that the primary component will fit the observed spectra very well, being identical to the single model, while the secondary component has a negligible contribution, with any value of $\vsini_2$ from available range. In the second case, the improvement factor will be small and thus these fits can be filtered out easily.
    \item Fitting a true binary system with a single star model: the single model will broaden its spectral lines to compensate for multiplicity in the spectrum by increasing a rotational velocity value. Thus, by using the selection criteria $\vsini_1+\vsini_2 < \vsini_0$,  SB2 candidates become easy to detect as they will likely be fitted with a large value of $\vsini_0$.
\end{enumerate}
Additionally, to filter out specious binary fits, criterion involving the improvement factor  ${\rm imp} > 0.1$ is used, following \cite{2022MNRAS.510.1515K}. This ensures that the binary model produces a significantly better fit to the observed spectra in comparison with the single star model. However we should note that our criteria will not select SB2s with fast rotating components if the difference  $|\rv_1 - \rv_2|$ is small, because in this case $\vsini_1+\vsini_2 > \vsini_0$. 

\section{Validation of criteria using synthetic stars}

To validate these selection criteria  two equal sets of 1000 mock single stars and 1000 mock binaries were created and used to apply the fitting technique and test the selection criteria for each cluster. Mock stars were created by selecting random points from the PARSEC \citep{2012MNRAS.427..127B} isochrone, which represent the positions of cluster's stars on the Hertzsprung-Russell with the same age but various masses. The isochrones are computed using the cluster ages and metallicities from Table~\ref{tab:sample}. The parameters used from the isochrone are mass, effective temperature, and surface gravity, which where selected within the ranges of the grid of synthetic spectral models.
For  mock binaries, values for $\rv_1$ and $\vsini_{1,2}$ are randomly selected from uniform distributions:  $U(\gamma - 100, \gamma + 100)~\kms$ and $U(5, 150)~\kms$ respectively, where $\gamma$ is the systemic velocity of the cluster. $\rv_2$ values are calculated using the formula:
\begin{equation}
    \rv_2 = \gamma(1 + q) - q \rv_1
    \label{eq: RVsecondary}
\end{equation}
 where the mass ratio was computed using masses from the isochrone.
\par
Mock single stars were generated by assigning the same values for both components for all parameters listed above, with $\rv_{0}=U(\gamma - 10, \gamma + 10)~\kms$ and $\vsini_{0}$ is randomly selected from the U(5, 150) $\kms$ distribution. \\

\begin{figure}
    \centering
    \includegraphics[width=\columnwidth]{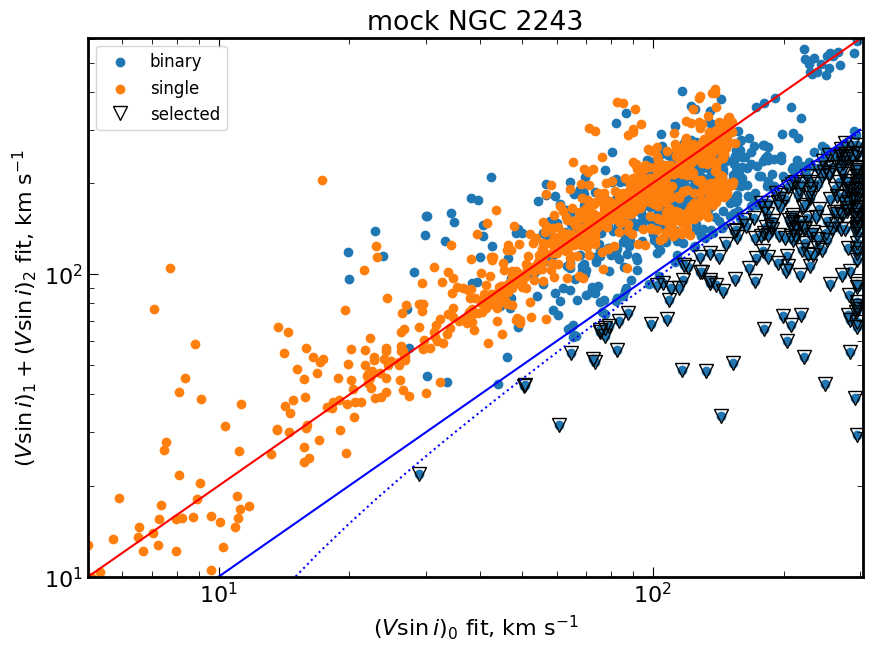}
    \includegraphics[width=\columnwidth]{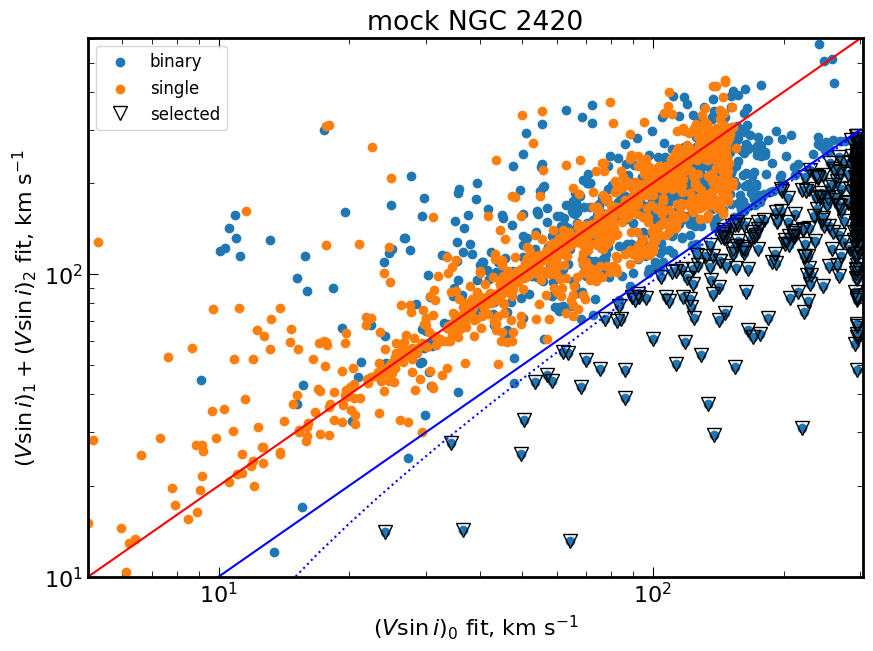}
    \includegraphics[width=\columnwidth]{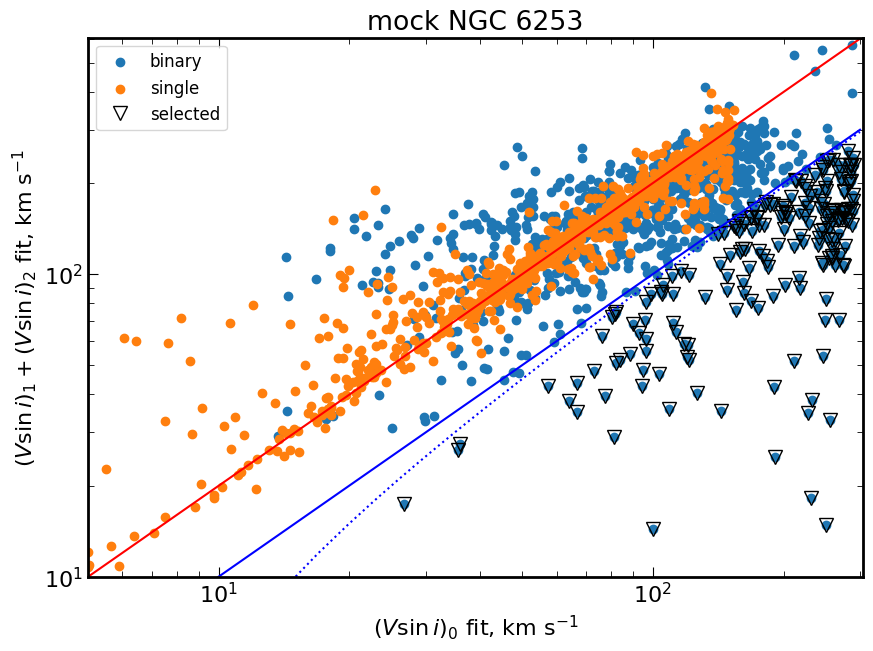}
    \caption{$\vsini_1 + \vsini_2$ vs. $\vsini_0$ plot for synthetic stars from three clusters with $\log{\rm Age(yrs)}>9$.  Blue and red lines shows relations $\vsini_1 + \vsini_2=\vsini_0$ and $\vsini_1 + \vsini_2=2 \vsini_0$ respectively. Dotted line shows relation $\vsini_1 + \vsini_2+5=\vsini_0$}
    \label{fig:synthetic1}
\end{figure}

\begin{figure}
    \centering
    \includegraphics[width=\columnwidth]{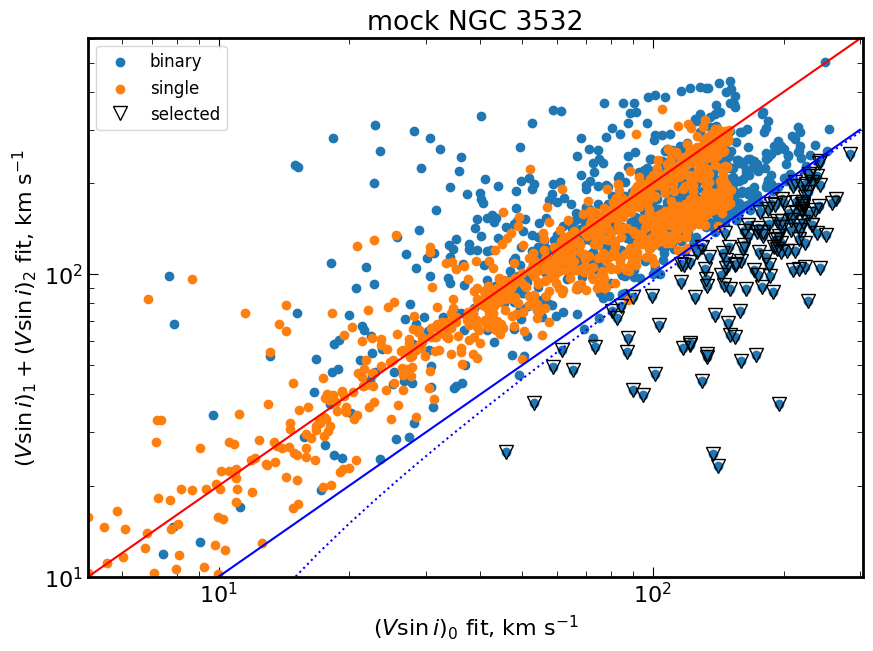}
    \includegraphics[width=\columnwidth]{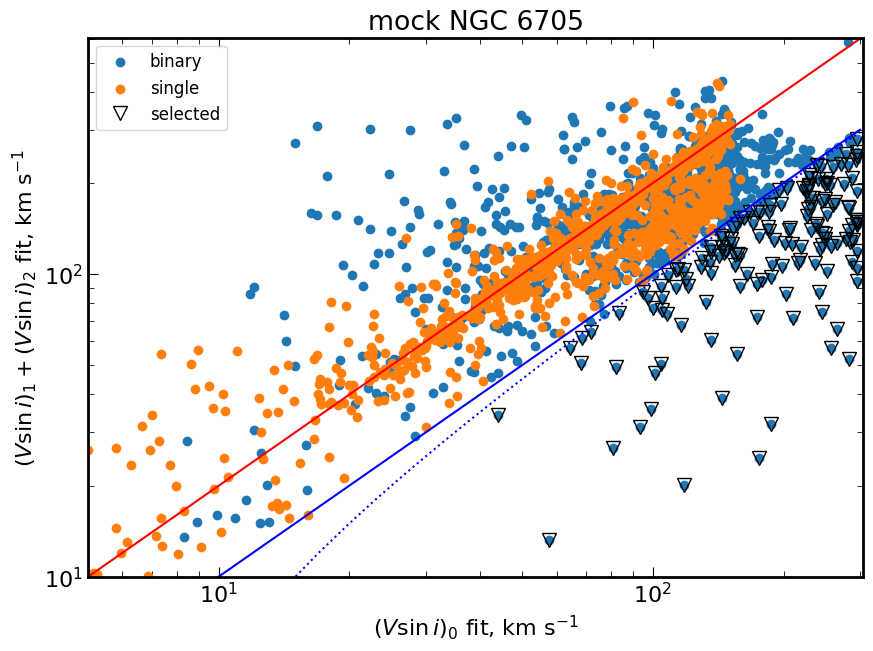}
    \caption{ Same as Fig.~\ref{fig:synthetic1} for synthetic stars from two clusters with $\log{\rm Age(yrs)}<9$.}
    \label{fig:synthetic2}
\end{figure}

Figures~\ref{fig:synthetic1},\ref{fig:synthetic2} display the results of the synthetic run. Solid lines represent the functions $\vsini_1+\vsini_2 = \vsini_0$ (blue) and $\vsini_1+\vsini_2 = 2 \vsini_0$ (red).  Blue doted line shows the relation $\vsini_1+\vsini_2 +5 = \vsini_0$. The value $\vsini_{\rm min}=5~\kms$ was set to prevent the selection of single stars, based of comparison between fitted and true values, where $\sigma(\Delta\vsini_0)<2.5~\kms$ for all clusters. The open triangles represent the stars that were selected by the criteria. Effective selection of binaries by the criteria would result in no false positives, i.e. no single stars would be selected as binaries.
It is apparent that no synthetic single stars (orange) appear below the blue dotted line and they strongly follow the relation $\vsini_1+\vsini_2 = 2 \vsini_0$ for reasons described in the previous section. Another overdensities formed by the single stars are visible slightly above the relation $\vsini_1+\vsini_2 =  \vsini_0$ for clusters NGC 2243, NGC 2420, NGC 3532 and NGC 6705. There we have single stars well fitted by the primary component, with negligible contribution from the secondary, which has small $\vsini_2$. Only for young cluster NGC 3532 some single stars appear below the blue line, but usage of $\vsini_{\rm min}=5~\kms$ filters them out.  
\par
These findings on mock stars demonstrate the reliability of this selection method for a given clusters. From the 1000 mock binaries in each cluster, the method is able to select  219, 291, 136, 191 and 155 for NGC 2243, NGC 2420, NGC 3532, NGC 6253 and NGC 6705 respectively. Although this indicates a relatively low success rate of 13-29\%, it reliably selects SB2 candidates without producing false positives; hence, the strength of this method lies in the purity of its selection,  not in completeness. Notice that success rate is smaller for younger clusters with higher $\feh$.

 \par 

Fitted parameters have good agreement with true parameters for a single star model, while for the binary model it can be worse. Usually it depends on radial velocity difference between the binary components and their contribution in total light \citep{2022MNRAS.510.1515K}. However we should note that for SB2 candidate identification purpose it not absolutely necessary to get exact parameters of both components with binary model, usually it is enough to see that binary model can fit spectrum much better than the single star model. Thus we can say that in our case binary model serves as a flexible ``composite spectrum template". To get more accurate estimation of the system's parameters one needs to fix mass ratio using some additional information \citep{2022MNRAS.510.1515K} or to use simultaneous fitting of several spectra, see \cite{tyc,j0647} for details.  

\section{Results and Discussion}
\begin{figure*}
    \includegraphics[width=0.79\textwidth]{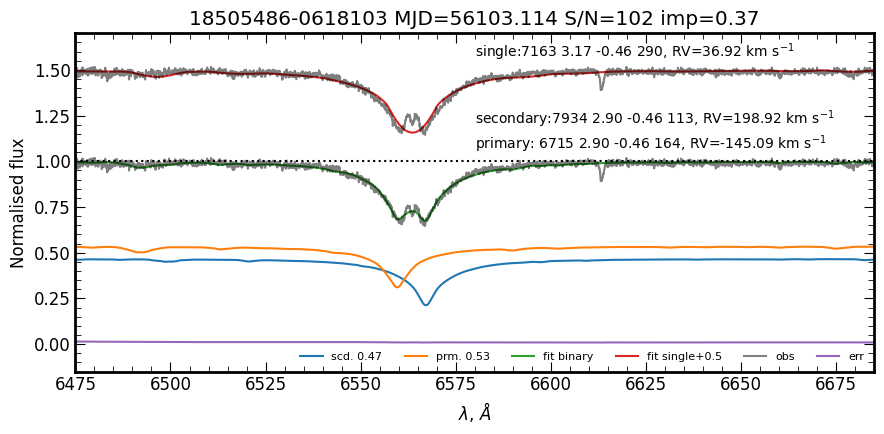}
        \caption{Examples of spectral fits with single-star and binary models for hot star with emission in $\ha$ line core.  This star was falsely selected as SB2 candidate, but can be easily eliminated thanks to the criterion on the large $\vsini_0$. Titles shows modified julian date (MJD), signal to noise ratio and improvement factor.  Best-fit spectral parameters ($\rv,~\teff,~\logg,~\feh,~\vsini$) are given for all fits.} 
        \label{fig:fake-emission}
\end{figure*}

We carefully inspected all resulting spectral fits by eye and found, that spectra with signal-to-noise ratio (S/N) $\snr<15$ are too noisy to make reliable selection. Additionally we found that for hot stars in cluster NGC~6705 with emission in $\ha$ line core binary model can provide fake result fitting it as a spectroscopic binary with very large $|\Delta\rv|>350~\kms$, see Figure~\ref{fig:fake-emission}. Fortunately such results can be easily filtered out as their value $\vsini_0$ hit the upper limit of the allowed range.  Also observations of many stars, especially in NGC~6705, have spectral line at $\lambda=6614$~\AA,~which is absent in our models. We suspect this is Diffuse Interstellar Band \citep{DIB}, as some clusters are located close to the Milky Way plane. Fortunately absence of this line in our spectral models have no effect to our selection as both single-star and binary model have been affected in similar way.
\par
\begin{figure}
     \centering
    \includegraphics[width=\columnwidth]{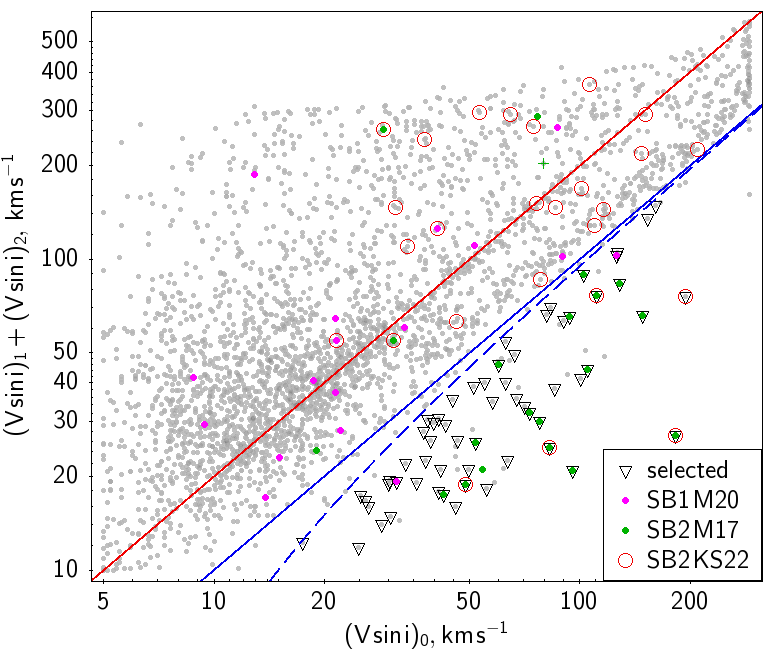}
    \caption{Same as Fig.~\ref{fig:synthetic1} for all observed spectra from five clusters. Selected SB2 candidates are shown as black open triangles. Known SB candidates are also highlighted.}
    \label{fig:obs}
\end{figure}

The criteria outlined above are used to select binary candidates after spectral analysis. In total 60 SB2 candidates are selected: two in NGC~2243, eight in NGC~2420 and NGC~3532, 17 in NGC 6253 and 25 in NGC~6705, see Tables~\ref{tab:final} and \ref{tab:full}.  Figure~\ref{fig:obs} shows the results of the selection. The selected SB2 candidates (black open triangles) fall below the blue dashed line as per the selection criteria. Known SB candidates from the literature are also highlighted. Among selected sample of  60 SB2 candidates, 18 were previously reported as spectroscopic binaries, which means that we identified 42 new ones. In comparison with our results for synthetic stars we have much smaller number of selected stars with big $\vsini_0$, which means that the radial velocity separation was smaller in comparison with mock binaries. Similarly to simulations, an overdensity parallel to the blue line (in the range 50 < $\vsini_0$ < 200 $\kms$) is also visible in our results. This is highly likely single stars. We found a relatively small number of SB2s in many clusters, which can be due to selection strategy of Gaia-ESO survey \citep{bragaglia}, which was not focused on the detection of spectral binaries \citep{2020A&A...635A.155M}. %
\par
Our selection finds only five SB2 candidates from 26 reported in \cite{2022MNRAS.510.1515K}, where selection was based on visual inspection of the plots and improvement factor. Many not selected SB2s contain two fast rotating stars, so our criterion can select them only for a very big radial velocity separation. 

\begin{figure*}
    \includegraphics[width=0.79\textwidth]{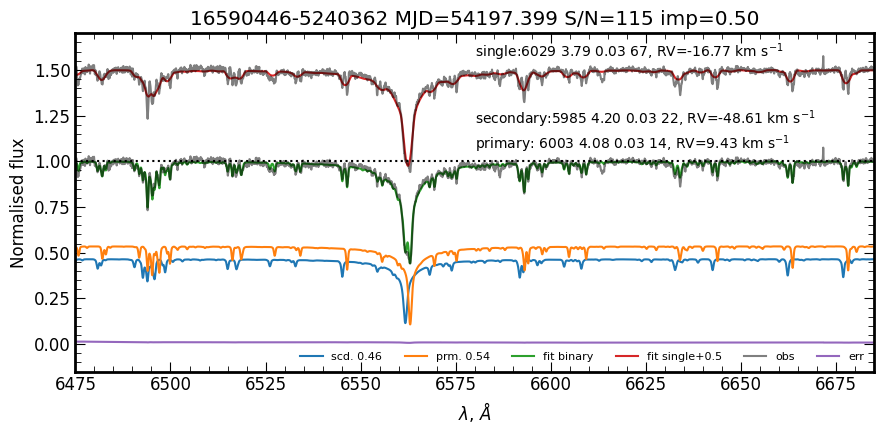}
    \includegraphics[width=0.79\textwidth]{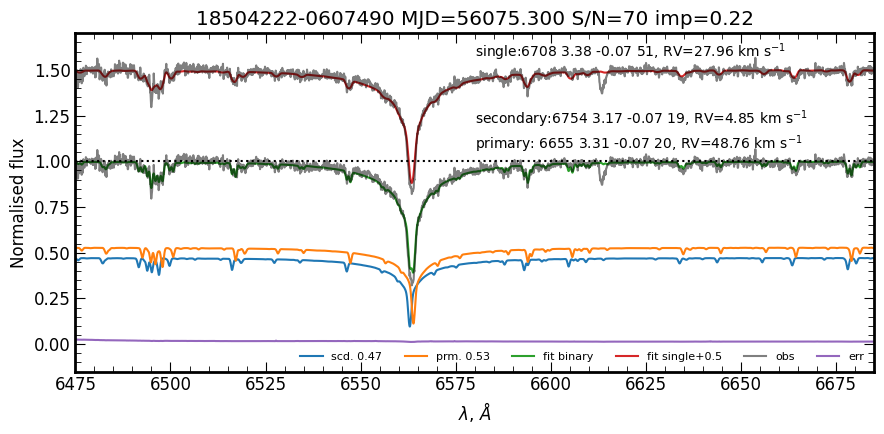}
    \includegraphics[width=0.79\textwidth]{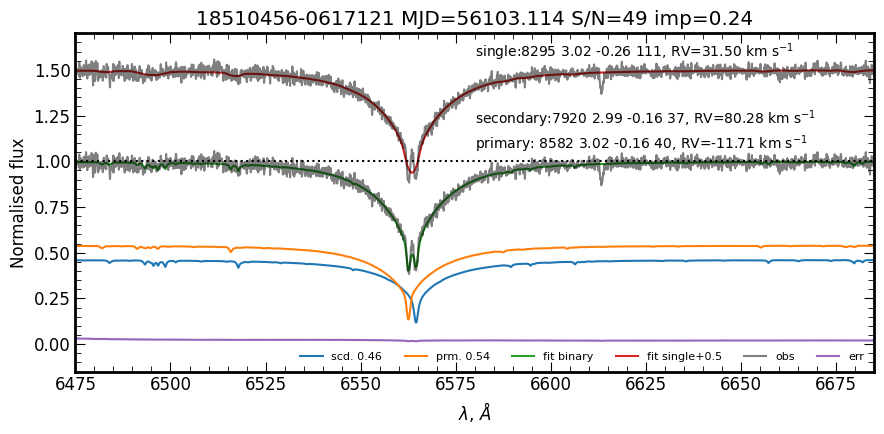}
        \caption{Same as Fig.~\ref{fig:fake-emission} for three SB2 candidates}
        \label{fig:obs_spectra}
\end{figure*}

Figure~\ref{fig:obs_spectra} shows examples of the fitting for three SB2 candidates. Visual inspection of the selected spectra confirms two spectral components, thus our criterion works well. However we should note, that all these SB2 candidates must be confirmed by additional spectral observations, as they can be result of chance alignment, which is more likely in stellar clusters due to crowding.

\begin{figure*}
    \includegraphics[width=0.79\textwidth]{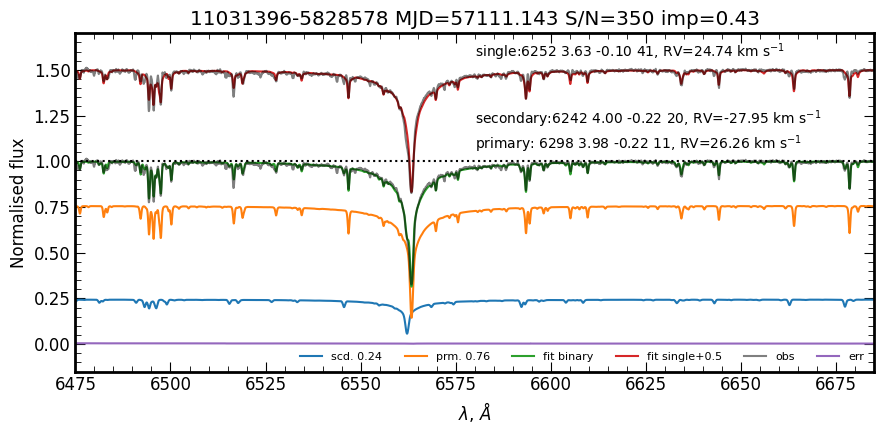}
    \includegraphics[width=0.79\textwidth]{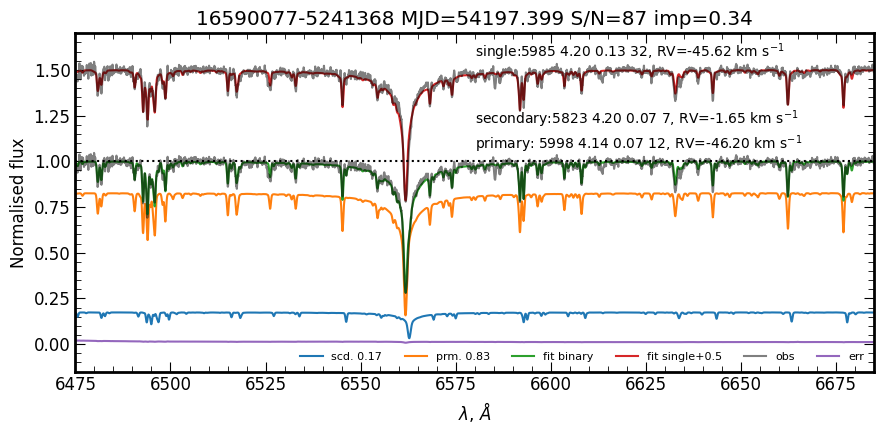}
    \includegraphics[width=0.79\textwidth]{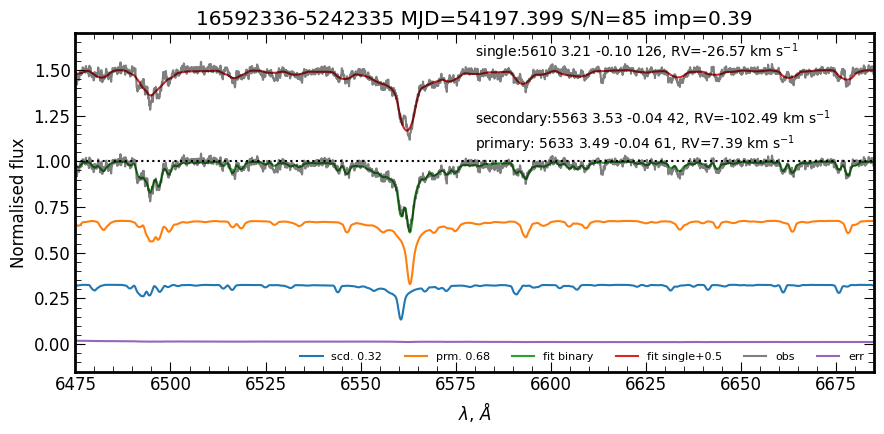}
        \caption{Same as Fig.~\ref{fig:fake-emission} for three SB2 candidates, previously reported as SB1.}
        \label{fig:sb1}
\end{figure*}

Three of the selected SB2 candidates, 16590077-5241368 and 16592336-5242335 from NGC~6253 and 11031396-5828578 from NGC~3532, are previously reported as SB1s in \cite{2020A&A...635A.155M} and non-single stars orbit catalogue \cite{gaia3}. Since two spectral components are evident in the spectra (see Figure~\ref{fig:sb1}), these systems should be reconsidered as SB2 candidates. 

\begin{figure*}
    \centering
    \includegraphics[width=0.79\textwidth]{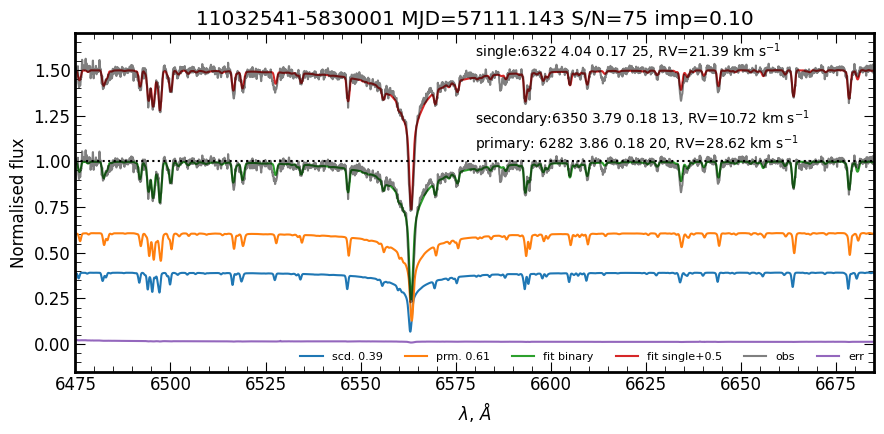}
    \includegraphics[width=0.79\textwidth]{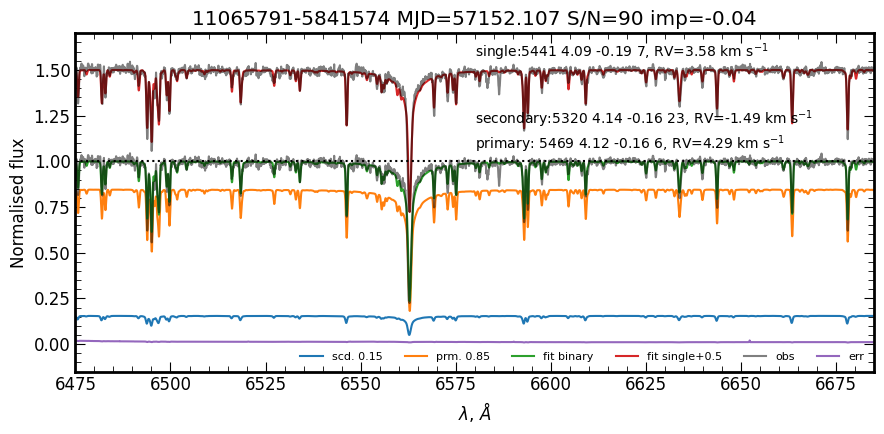}
    \caption{Same as Fig.~\ref{fig:fake-emission} for two known eclipsing binaries from NGC 3532: HD 96609 (top) and HD 303734 (bottom), but not selected as SB2 by our method because the binary models do not significantly improved the fit over the single star models.}
    \label{fig:eclipsing}
\end{figure*}

The two eclipsing binaries in NGC 3532, HD 96609 (11065791-5841574) and HD 303734 (11032541-5830001), described by \cite{2022MNRAS.509.1912O}, were included in the present analysis. However, their spectra were not selected by the method as SB2 candidates. The spectra, depicted in Figure~\ref{fig:eclipsing}, show that the binary fit does not significantly improve upon the single star model. It is possible that these spectra were observed, when the stars were in the part of their orbits where they had no radial velocity discrepancy.\\

 The implementation of this method has found several limitations. First, stars with temperatures that are lower or higher than the range of the grid used in the model lead to high errors in the spectral fitting. Specifically, the model could not account for spectral lines profiles of the hot stars ($\teff>8700$ K) that exceeded the range of the grid, thus it could have attempted to corrected for broadening by increasing the estimated rotational velocity $\vsini_0$, which could lead to erroneous selection of single stars as SB2s. Also, very cool stars of temperatures $\teff<5000$ K are poorly fitted, as they can have molecular bands, absent in our models. Such fits were excluded after the visual inspection. Low S/N of the observed spectra was found to be another source of false selection. The binary components of the model fit to the noise of the spectra, producing two spectral components that may not truly appear in the observed spectra.


\section{Conclusions}
This paper tests the reliability of the selection criteria proposed by \cite{2022MNRAS.517..356K} to detect SB2s by applying them to five open clusters. The method for selection depends on the assumption, that a single star model compensates for the disparity of double spectral components by increasing the estimated projected rotational velocity.  Analysis of  both synthetic and observed datasets shows that this method is effective at selecting binaries without producing false positive selections. The method found 60 SB2 candidates overall, with 18 of them previously documented in literature. We plan to apply this selection for all available HR15N spectra in the future paper.

\section*{Acknowledgements}
 We are grateful to the anonymous referee for a constructive report and suggestions, which greatly improved the paper. 
MK is grateful to his parents, Yuri Kovalev and Yulia Kovaleva, for their full support in making this research possible. 
This work is supported by National Key R\&D Program of China (Grant No. 2021YFA1600401/3), International Centre of Supernovae, Yunnan Key Laboratory (No. 202302AN360001) and by the Natural Science Foundation of China (Nos. 12090040/3, 12125303, 11733008).
The authors gratefully acknowledge the “PHOENIX Supercomputing Platform” jointly operated by the Binary Population Synthesis Group and the Stellar Astrophysics Group at Yunnan Observatories, Chinese Academy of Sciences. 
Based on data products from observations made with ESO Telescopes at the La Silla Paranal Observatory under run IDs 079.D-0825, 188.B-3002 and 193.B-0936.
This research has made use of NASA’s Astrophysics Data System, the SIMBAD data base, and the VizieR catalogue access tool, operated at CDS, Strasbourg, France. It also made use of TOPCAT, an interactive graphical viewer and editor for tabular data \citep[][]{topcat}.

\section*{Data Availability}
The data underlying this article will be shared on reasonable request to the corresponding author.



\bibliographystyle{mnras}





\appendix

\section{Tables with selected SB2 candidates}

Table~\ref{tab:final} contains a list of selected SB2s and their object names as recorded in the ESO archive together with Gaia DR3 \texttt{source\_id}.
Table~\ref{tab:full} have some additional info, like best-fit spectral parameters for the single star and the binary models. These parameters should be used with care, because they are serving just as a ``best matching template". 

\begin{table*}
    \centering
    \caption{Catalogue of selected SB2 candidates. M+17 - \protect\cite{2017A&A...608A..95M}, M+20 - \protect\cite{2020A&A...635A.155M}, KS22 - \protect\cite{2022MNRAS.510.1515K}, GDR3 - \protect\cite{gaia3}.}
    \begin{tabular}{lcccc}
\hline
\hline
Cluster & ESO cname & GDR3 \texttt{source\_id} & $G$, mag& remark\\
\hline

   NGC~2243 & 06290412-3114343 & 2893945772888126720 & 14.934553& SB2 M+17\\
   NGC~2243 & 06292643-3115445 & 2893942921030310912 & 16.304037\\
   NGC~2420 & 07374440+2136232 & 865412923480423680 & 17.150282\\
   NGC~2420 & 07380618+2137236 & 865402860372744704 & 15.4575\\
   NGC~2420 & 07382246+2132287 & 865399939794989952 & 15.667797\\
   NGC~2420 & 07382338+2131396 & 865396950497755264 & 16.9582\\
   NGC~2420 & 07382366+2133430 & 865401485983205504 & 15.800236\\
   NGC~2420 & 07383323+2135213 & 865401657781870080 & 15.149413\\
   NGC~2420 & 07384133+2134225 & 865398939066030976 & 16.65745\\
   NGC~2420 & 07384553+2129385 & 865397122296423424 & 14.6743\\
   NGC~3532 & 11020707-5855342 & 5338686027293771904 & 17.397034\\
   NGC~3532 & 11022934-5845476 & 5338682934919078784 & 17.219435\\
   NGC~3532 & 11024689-5844177 & 5338706299537601280 & 11.168904\\
   NGC~3532 & 11030296-5900435 & 5338675758080306816 & 12.7284046& astrometric acceleration GDR3\\
   NGC~3532 & 11031396-5828578 & 5338715782825370368 & 12.163282& SB1 GDR3\\
   NGC~3532 & 11054586-5857236 & 5338651225224549632 & 17.295105\\
   NGC~3532 & 11065099-5838199 & 5340161576236734848 & 11.287692\\
   NGC~3532 & 11073165-5848233 & 5340147729260314752 & 15.834448\\
   NGC~6253 & 16583965-5235499 & 5935994452129297152 & 14.892061\\
   NGC~6253 & 16584301-5247576 & 5935967136135123840 & 14.236232\\
   NGC~6253 & 16585321-5242431 & 5935991291033271168 & 14.035797\\
   NGC~6253 & 16585749-5242259 & 5935944389986065408 & 14.101318\\
   NGC~6253 & 16590035-5245080 & 5935944183827814016 & 14.592207\\
   NGC~6253 & 16590077-5241368 & 5935945897460206080 & 14.928719& SB1 M+20\\
   NGC~6253 & 16590417-5237044 & 5935993524416300800 & 16.764376\\
   NGC~6253 & 16590446-5240362 & 5935946039253519232 & 14.361807\\
   NGC~6253 & 16590609-5242566 & 5935945798735308160 & 14.448471\\
   NGC~6253 & 16590624-5242009 & 5935945833095064192 & 14.367528\\
   NGC~6253 & 16590845-5240439 & 5935945970534033408 & 14.527414\\
   NGC~6253 & 16591059-5242177 & 5935945936174263040 & 14.72464\\
   NGC~6253 & 16591586-5245025 & 5935944046388593408 & 14.932591\\
   NGC~6253 & 16591656-5244053 & 5935945523857356800 & 14.223827\\
   NGC~6253 & 16591716-5242442 & 5935945661296332032 & 14.544811\\
   NGC~6253 & 16592297-5235100 & 5935952735057641856 & 16.270937\\
   NGC~6253 & 16592336-5242335 & 5935945695656057216 & 14.746293& SB1 M+20\\
   NGC~6705 & 18502379-0615209 & 4252509585784232576 & 16.581614\\
   NGC~6705 & 18503230-0617112 & 4252509246592782464 & 13.291144& SB2 M+17\\
   NGC~6705 & 18503690-0621100 & 4252496636566974592 & 16.33684& SB2 M+17\\
   NGC~6705 & 18503840-0617048 & 4252497525515917440 & 16.362226& SB2 M+17\\
   NGC~6705 & 18504183-0623464 & 4252495502695435776 & 15.965503\\
   NGC~6705 & 18504188-0622567 & 4252495605774692096 & 16.201315\\
   NGC~6705 & 18504222-0607490 & 4252518798601477504 & 15.961194\\
   NGC~6705 & 18504649-0611443 & 4252516118541710336 & 15.716564& SB2 M+17\\
   NGC~6705 & 18504896-0614324 & 4252503955192670080 & 15.743275\\
   NGC~6705 & 18505045-0617310 & 4252503096165723776 & 18.395308\\
   NGC~6705 & 18505561-0614552 & 4252503920832494464 & 16.174961& SB2 M+17\\
   NGC~6705 & 18505701-0615044 & 4252503710273899264 & 15.850019\\
   NGC~6705 & 18505726-0609408 & 4252517767809140864 & 16.413301\\
   NGC~6705 & 18505933-0622051 & 4252493131873385600 & 16.42047& SB2 M+17\\
   NGC~6705 & 18510072-0609118 & 4252517802168868352 & 16.614513& SB2 M+17\\
   NGC~6705 & 18510286-0615250 & 4252503744633664640 & 13.157031& SB2 M+17\\
   NGC~6705 & 18510288-0617400 & 4252502786927640960 & 17.8157\\
   NGC~6705 & 18510456-0617121 & 4252502782561000704 & 12.734941& SB2 M+17\\
   NGC~6705 & 18511131-0617491 & 4252502546409177856 & 17.137434\\
   NGC~6705 & 18511134-0616106 & 4252502919999990528 & 13.862979& SB2 M+17\\
   NGC~6705 & 18511220-0617467 & 4252502542042872192 & 15.212772& SB2 M+17\\
   NGC~6705 & 18511434-0617090 & 4252502645122095232 & 16.732573& SB2 M+17\\
   NGC~6705 & 18512031-0609011 & 4252506218523086080 & 12.372231& SB2 KS22\\
   NGC~6705 & 18512604-0610504 & 4252505943652338432 & 15.581753\\
   NGC~6705 & 18513193-0612518 & 4252504814185976960 & 16.381422& SB2 M+17\\
    
\hline
\end{tabular}


    \label{tab:final}
\end{table*}

\begin{table}
    \centering
    \caption{Additional information for selected SB2 candidates. Full table is available online as a supplemental material.}
    \begin{tabular}{lc}
\hline
parameter & unit\\
\hline
ESO cname\\
${\teff}_0$ & K\\
$\logg_0$ & dex\\
$\feh_0$ & dex\\
$\vsini_0$ & $\kms$\\
${\teff}_2$ & K\\
$\logg_2$ & dex\\
$\feh_{1,2}$ & dex\\
$\vsini_2$ & $\kms$\\
${\teff}_1$ & K\\
$\logg_1$ & dex\\
$\vsini_1$ & $\kms$\\
$\rv_0$ & $\kms$\\
$\rv_2$ & $\kms$\\
$\rv_1$ & $\kms$\\
$q$ \\
{\rm imp} \\
$\snr$ & ${\rm pix}^{-1}$\\
MJD & d\\
GDR3 \texttt{ source id}\\
$G$& mag\\
\hline
\end{tabular}
    \label{tab:full}
\end{table}





\bsp	
\label{lastpage}
\end{document}